# Wiring surface loss of a superconducting transmon qubit

**Nikita S. Smirnov,[1,2] Elizaveta A. Krivko,[1] Anastasiya A. Solovieva,[1] Anton I. Ivanov,[1] and Ilya A. Rodionov[1,2*]**

[1]FMN Laboratory, Bauman Moscow State Technical University, Moscow, 105005, Russia

[2]Dukhov Automatics Research Institute, VNIIA, Moscow, 127030, Russia

* irodionov@bmstu.ru

## Abstract

Quantum processors using superconducting qubits suffer from dielectric loss leading to noise and dissipation. Qubits are usually designed as large capacitor pads connected to a non-linear Josephson junction (or SQUID) by a superconducting thin metal wiring. Here, we report on finite-element simulation and experimental results confirming that more than 50% of surface loss in transmon qubits can originated from Josephson junctions wiring and can limit qubit relaxation time. Extracting dielectric loss tangents capacitor pads and wiring based on their participation ratios, we show dominant surface loss of wiring can occur for real qubits designs. Then, we simulate a qubit coupled to a bath of individual TLS defects and show that only a small fraction (∼18%) of coupled defects is located within the wiring interfaces, however, their coupling strength is much higher due to stronger electromagnetic field. Finally, we fabricate six tunable floating transmon qubits and experimentally demonstrate up to 20% improvement in qubit quality factor by wiring design optimization.

## INTRODUCTION

Quantum processors and simulators comprising tens or even hundreds superconducting qubits have recently been demonstrated[1–5]. Quantum gates errors hinder further size and complexity growth of superconducting circuits and quantum algorithms. On the one hand, reducing two-qubit gate errors to less than 0.1% opens a practical way to implement quantum error correction codes[6,7]. Meanwhile, quantum error correction is mathematically difficult and requires enormous qubit resources. On the other hand, with a reduced gate errors a useful quantum advantage can be achieved near term using special task quantum algorithms and error mitigation[1,8]. But superconducting quantum bits have natural internal sources of noise and decoherence limiting quantum gates fidelity.

A large part of qubits loss is due to microscopic tunneling defects, which form parasitic two-level quantum systems (TLS)[9,10] and resonantly absorb electric energy from the qubit mode dissipating it into phonons or quasiparticle bath[11–14]. It is well-known, that such defects reside in the interfaces and surface native oxides around qubit electrodes: metal-substrate (MS), substrate-air (SA), metal-air (MA)[15–19]. This source of qubit loss could be mitigated by reducing the amounts of lossy dielectrics (minimizing Josephson junction area[20,21], using better materials and defect-free fabrication techniques[22–24]). Another approach for loss mitigation is increasing qubits footprint[25–27] by minimizing an electric field in the interfaces and preventing TLS excitation due to coupling with their dipole moment.

Qubit relaxation caused by dielectric losses could be decomposed into participations from each material and qubit components:

$$\frac{1}{T_1} = \frac{\omega}{Q} = \omega \sum_i p_i \tan\delta_i, \qquad (1)$$

where $T_1$, $\omega$ and $Q$ are the relaxation time, angular frequency and quality factor of the qubit, $\tan\delta_i$ is the dielectric loss tangent of the $i^{th}$ material or component, and $p_i$ is their participation ratio defined as the fraction of electric field energy stored within inside this material or component.

One can imagine a superconducting transmon qubit[28] as a non-linear LC oscillator, where the Josephson junction or SQUID define a non-linear inductance and the superconducting metal pads define a capacitor. Josephson junction or SQUID loop electrodes have to be electrically connected to the capacitor pads. Such a connection is commonly realized as a thin metal wire, which we call **leads** in this paper. The design of the frequency tunable two-padded floating transmon qubit which is investigated in this study is shown in Fig. 1a. Usually, in order to improve qubit relaxation time (dilute an electromagnetic field and lower interfaces participation ratio), the gap (G, Fig. 1a) between the capacitor pads is increased. However, it requires an increased length of the Josephson junction connecting wires. Moreover, in case of a flux-tunable qubit the wiring length becomes even longer to form a SQUID loop and move it closer to the gap edge and flux-control line. Figure 1b demonstrates qubit pads and qubit wiring (leads and SQUID loop) participation ratios versus qubit gap width. One can notice, as the gap width increases the capacitor pads participation ratio decreases, but the leads with SQUID participation ratios increase. When the gap width is more than 110 μm, then the participation ratio of the leads with SQUID become dominant and further gap widening is impractical. Thus, a relaxation time of a properly designed qubit is limited by the leads and SQUID loss, if their loss tangent is comparable to the capacitor pads one. To further increase the qubit relaxation time, we optimize the leads width as illustrated in Fig. 1c.

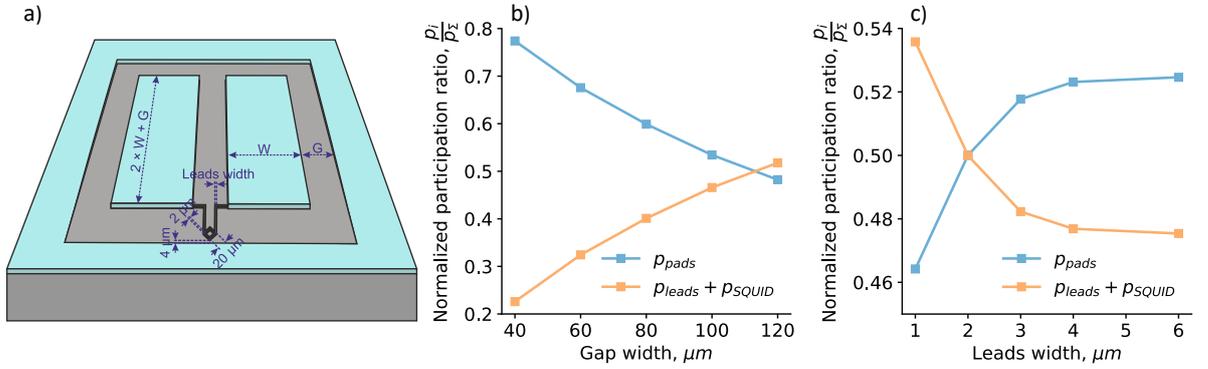

**Fig. 1 Flux-tunable floating transmon qubit participation ratios. a** Sketch of transmon qubit with a standard wiring and SQUID loop (black) close to control line. Capacitor pad dimensions and gap width (G) are proportional to make qubit symmetrical in both directions. **b** Normalized participation ratio of the pads (blue dots) and wiring (orange dots, include leads and SQUID loop) vs. gap width (G). For each gap width capacitor pad width is adjusted to make qubit charge energy $E_C$ equal to 220 MHz. Leads width is 2.5 μm. **c** Normalized participation ratios vs. leads width. Gap width (G) and pad width (W) are chosen equal to 120 μm and 204 μm, correspondingly. Curves flatten out when the participation ratio of the SQUID loop in wiring becomes dominant.

We calculated the participation ratios using a similar method as in Ref.[25], but with modifications to be able to analyze asymmetrically located SQUID loop with long leads (see Supplementary for details). When we calculate the participation ratios of qubit components, we consider bulk superconductor and crystalline dielectric to be lossless ($tan\delta_i < 10^{-6}$ for silicon[29] and sapphire[30]), so that:

$$p_{pads} = p_{pads_{MA}} + p_{pads_{SA}} + p_{pads_{MS}} \quad (2)$$
$$p_{SQUID} = p_{SQUID_{MA}} + p_{SQUID_{SA}} + p_{SQUID_{MS}} \quad (3)$$
$$p_{leads} = p_{leads_{MA}} + p_{leads_{SA}} + p_{leads_{MS}} \quad (4)$$

In order to mitigate surface dielectric loss in qubits, previous work are primarily focused on capacitor pads design modifications[25–27,31], investigation of Josephson junction contribution with so-called bandages and fabrication improvements[32–36]. Despite the remarkable results achieved in these works, contribution from the Josephson junction wiring has mainly remained ignored with rare exceptions. In Ref.[25] there were extracted participation ratios for junction leads, but only for 3D cavity qubits with a single Josephson junction. Recent study [37] has analytically predicted, that a significant fraction of surface loss comes from the wiring that connects the qubit to the capacitor. In this work, we experimentally studied the contribution of leads and SQUID to the overall qubit surface dielectric loss. We performed finite-element electromagnetic simulations of the transmon qubits with different leads geometries in order to analyze their contribution to the surface losses. Then, we experimentally measured qubits relaxation times and extracted the dielectric loss tangents of the qubit elements associated with the capacitors, leads and SQUID. We also compared two different methods for leads fabrication: etch and lift-off. Then, we demonstrated good agreement between the measured qubit quality factors and the proposed model, so it could be used to further qubit design optimization reducing surface dielectric losses. The dielectric loss model and loss tangents extracted in this study can be applied to improve perspective superconducting qubit, e.g. fluxoniums[38].

## RESULTS AND DISCUSSION

To analyze the contribution of capacitor pads, leads and SQUID in the total qubit loss, we fabricated six tunable floating transmon qubits on the same chip, so we can assume the same loss tangents of each interface for all the qubits. The fabricated qubits have the same design, except different wiring geometry, as shown in Figure 2a. They were designed to accentuate participation in the Josephson junction leads. Therefore, we refer to the designs as "long leads", "regular leads" and "wide leads". The calculations of participation ratios were performed considering 3 nm thick dielectric interface layers MA, SA and MS with fixed dielectric constant $\epsilon = 10$. Table 1 summarizes the parameters of the fabricated qubits.

**Table 1. Parameters of fabricated qubits.**

| Qubit# (lift-off/etch) | Design | $p_{pads}$ | $p_{leads}$ | $p_{SQUID}$ | $f_{q,max}$, (GHz) | $\eta_q$, (MHz) |
|---|---|---|---|---|---|---|
| 1/6 | long | 1.852 | 3.312 | 0.613 | 4.825/4.739 | 179 |
| 3/4 | regular | 1.938 | 1.247 | 0.694 | 4.999/5.164 | 215 |
| 5/2 | wide | 2.086 | 0.652 | 0.724 | 4.824/5.152 | 182 |

Qubit numbers correspond to the numbers on Fig. 2b. Etch and lift-off are methods of leads fabrication. Participation ratios of the elements $p_i$ are multiplied by $10^4$. $f_{q,max}$ and $\eta_q$ are the experimentally measured qubit "sweet spot" frequency and anharmonicity.

Qubit capacitor pads electrodes and ground plane were wet-etched in 120 nm Al film. The Josephson junctions were defined using electron beam lithography and shadow-mask evaporation. Then, Al bandages were deposited in order to either short the stray junctions or connect the leads with the capacitor pads. Finally, superconducting airbridges were fabricated to suppress the parasitic microwave modes. See Methods for more fabrication details.

In order to estimate the effect of fabrication process to the leads surface losses, we compared two methods of leads patterning. Three qubits on the chip have the leads fabricated together with the SQUID loop using lift-off process. The leads of the other three qubits were patterned using optical lithography and, then, wet-etched together with the capacitor pads. A scanning electron microscope (SEM) image of the sample with six floating transmon qubits is shown in Fig. 2b.

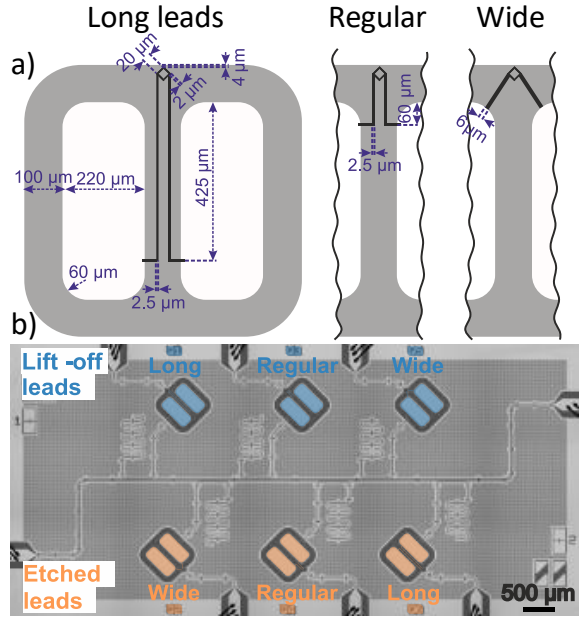

**Fig. 2 Tunable floating transmon qubit designs and fabricated sample. a** Designs of three transmon qubits with "long leads", "regular leads" and "wide leads". Dimensions of square SQUID loop and capacitor pads are fixed for all three designs. Gaps are shown in grey and capacitor pads with the surrounding ground plane in white. **b** False-colored SEM image of the sample featuring six qubits coupled to readout resonators. Qubits are initialized and readout is performed via the single feedline. The qubits with etched Josephson junction leads are colored in orange and with lift-off leads in blue.

All the qubits are individually coupled to $\lambda/4$ resonators with different frequencies ranging from 6 up to 6.45 GHz for dispersive readout. State-dependent dispersive shifts, qubit-resonator detuning and resonator widths are designed to both distinguish the readout signals and push the Purcell-limited relaxation time as high as possible (> 1 ms), so it does not affect the qubits relaxation time ($T_1$). See Methods for more experimental setup details.

In this work we spectrally resolve $T_1$ of the frequency-tunable floating transmons. We recalculate $T_1$ in quality factors (Q-factor) and determine their mean value and confidence interval for each qubit. The experimental pulse sequence we used to measure $T_1$ at a single frequency is the follows: the qubit is initialized in the $|1\rangle$ state by a microwave drive pulse, flux-tuned to the frequency of interest, where we wait for a varied delay time, and then measure the qubit state. To obtain one $T_1$ curve, we repeat this experiment for 21 equally-spaced delays with 4000 shots each in the range 10 to 400 μs. Using this sequence, we swept the qubit frequency over a range of at least 300 MHz with 1 MHz step. It took us approximately 6 hours to measure the entire 300 MHz spectrum. Qubit's spectra, converted into quality factors, and distributions are presented in Fig. 3. We notice that qubit spectra have Lorentzian-like regions with strong relaxation (red dots in Fig.3a and in Fig.3b). We attributed these resonances to the modes of qubit flux control lines or microwave package cavity modes, as their shape maintained after repeated cooldowns. These peaks can be successfully suppressed with IR eccosorb-filters in flux control cables[39]. However, these filters may slightly limit $T_1$, that is why in this study we don't use IR filtering for the flux control cables. We excluded Lorentzian regions from the analysis, as they are not connected with surface dielectric loss.

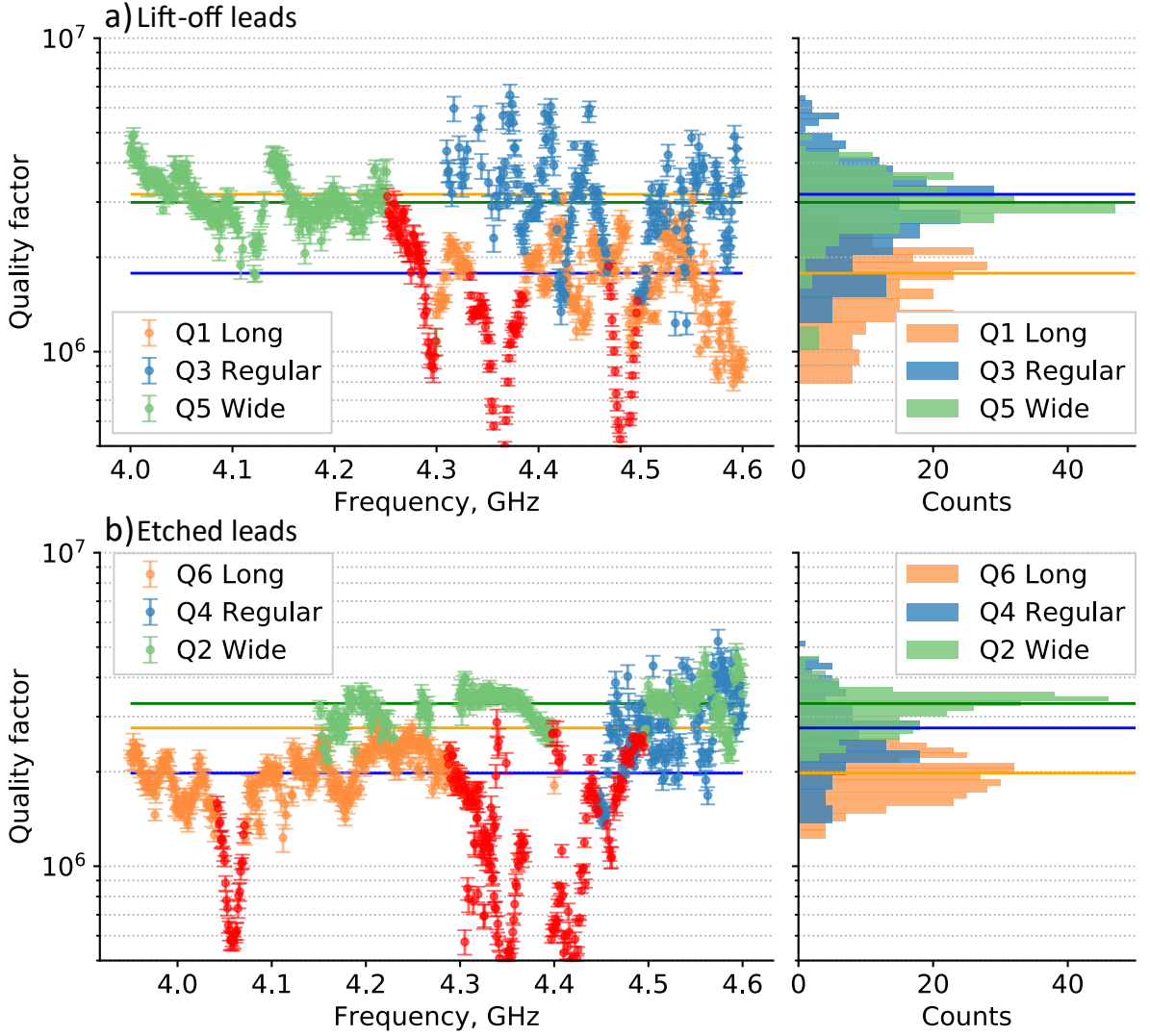

**Fig. 3 Measured qubits quality factors vs. qubit frequency.** Orange, blue and green dots correspond to the "long leads", "regular leads" and "wide leads", respectively. Error bars are quality factor fit errors. All the dots with big errors (> 10%) are excluded from the plot. Parasitic modes dots are colored in red and are excluded from the distribution plots. Solid horizontal lines show the median quality factor values. **a** Qubits with lift-off leads. Median Q-factors are $Q_{Q1} = 1.78 \times 10^6$, $Q_{Q3} = 3.17 \times 10^6$, $Q_{Q5} = 2.99 \times 10^6$. **b** Qubits with etched leads. Median Q-factors are $Q_{Q6} = 1.98 \times 10^6$, $Q_{Q4} = 2.76 \times 10^6$, $Q_{Q2} = 3.30 \times 10^6$.

We compared the qubit components loss tangents extracted with surface loss extraction (SLE)[19] process of lift-off and wet etch leads (see Table 2). Qubit quality factor, written through Eq. 1, can be represented in matrix form as a function of participation matrix $[P]$ (number of columns equal to number of qubit components, including pads, leads, SQUID; number of rows equal to number of designs) and loss tangent vector $[\tan\delta]$ (number of columns equal to the number of qubit components):

$$[1/Q] = [P][\tan\delta] \qquad (6)$$

In order to determine the uncertainty of the extracted loss tangents, we perform Monte Carlo simulations. We sampled 10000 Q-factors using mean values and standard deviations taken from the experimental data. Then, taking each sampled Q-factors and calculated participation ratio in the matrix $[P]$, we found the least square solution for the loss tangents using Eq. 6. Based on the extracted loss tangents and calculated participation ratios in Eq.1, we determined the predicted Q-factors, shown in the Fig. 4a as dashed lines. Dark blue dots with solid lines (Fig. 4a) show median values of the measured Q-factors versus leads design for lift-off leads (orange dots with solid lines for etched leads). The horizontal and the vertical errors bars correspond to the standard deviation error of the measured and simulated Q respectively. One can notice quality factors improvement as Josephson junction leads participation ratio decreases. This trend shows that leads with a high participation ratio suppress significantly the qubit quality factor. Qubits with "wide" shortened leads have a higher median quality factor ($Q_{Q2}$ = $3.30 \times 10^6$, $Q_{Q5}$ = $2.99 \times 10^6$) compared to qubits with narrower and longer "regular" leads ($Q_{Q3}$ = $3.17 \times 10^6$, $Q_{Q4}$ = $2.76 \times 10^6$), the worst case is "long" lead ($Q_{Q1}$ = $1.78 \times 10^6$, $Q_{Q6}$ = $1.98 \times 10^6$). The only exception in this experiment is the qubit (Q3) with lift-off "regular" leads, which median quality factor become a slightly higher than for qubit (Q5) with "wide" leads. We assume this as a statistical error, as the qubit (Q3) has the worst scatter in $T_1$, quality factor measurements (probably, due to dynamics of strongly coupled TLS [40,41]). Moreover, the predicted quality factor curves for lift-off and etched leads have the same shape and correspond well to measured curve for etched leads, which may also indicate a statistical error for qubit (Q3) measurements.

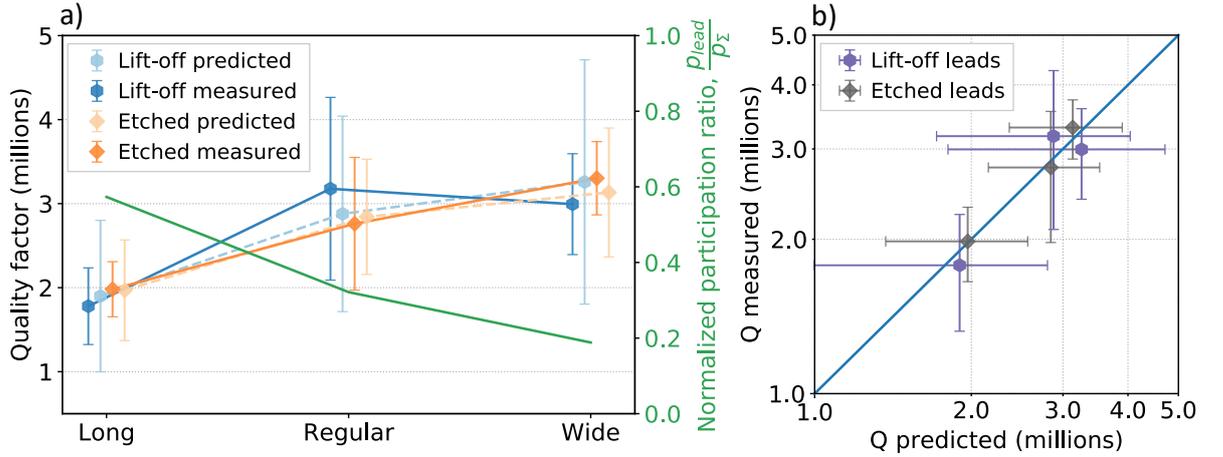

**Fig. 3 Measured and predicted qubits quality factors. a** Qubits quality factors and normalized participation ratio as a function of leads geometry. Each data point represents median quality factor obtained from measured qubit $T_1$ spectra (solid lines) or sampled quality factors (dashed lines). Dark blue and dark orange points represent quality factors for measured qubits with lift-off and etched leads, respectively. Light blue and light orange points represent predicted quality factors for lift-off and etched leads, respectively. Plotted normalized participation ratio of leads (solid green line) shows that leads have non-negligible effect. **b** Predicted quality factors compared to measured quality factors of qubits with etched (dark gray) and lift-off (purple) leads. The blue line represents perfect agreement between the measured and predicted quality factors. All the error bars on the plots correspond to 68% confidence interval.

We attribute the uncertainty of the extracted loss tangents to several sources. First, the scatter in qubit's quality factor measurement statistics results in wider range of possible linear equations solutions. Second, although we strived to maximize the participation ratio of Josephson junction leads, we were not able to make a design with the participation of the specific component totally prevailing over the other qubit components. The maximum participation ratio of Josephson junction leads relative to the total ratio of all the qubit components in our set of designs was 0.364. As we expected, the loss tangents of capacitive pads and SQUIDs for qubits with lift-off and wet etched leads are almost the same, since they were fabricated on the same chip. The loss tangent of etched leads is smaller than lift-off one. We attribute this to a degraded MS interface due to e-beam resist residuals[42] and a worse MA interface, which was additionally passivated after shadow deposition (a thicker amorphous oxide can occur). We also note that the cross-section of shadow-evaporated structures is quite complex in practice. There are both exposed oxidized areas of bottom electrode and areas of top electrode metal covering the bottom one. This feature distorts the cross-section of lift-off wiring and SQUID, making them different from a rectangle, which introduces additional MA interfaces. As it is quite complicated, we do not take it into account in our model for participation ratios simulation, that could affect the results accuracy. One can see, that loss tangent of the SQUID loops is lower than the leads for both fabrication routes. The SQUID loops have much smaller footprint area, but still non-negligible TLS defect (see our TLS spectrum analysis below).

**Table 2. Experimentally extracted loss tangents of qubit elements.**

| Process | $\tan\delta_{pads}$ (× $10^4$) | $\tan\delta_{leads}$ (× $10^4$) | $\tan\delta_{SQUID}$ (× $10^4$) |
|---|---|---|---|
| Lift-off | 10.4 ± 3.9 | 9.2 ± 4.5 | 3.7 ± 2.1 |
| Wet etch | 11.3 ± 3.4 | 7.9 ± 3.8 | 4.0 ± 1.7 |

Here we show the values obtained using SLE process. It represents different leads patterning processes.

Surface losses mainly arise from the interaction of qubits with a bath of incoherent TLS defects. The relaxation rate depends on the number of defects coupled to the qubit mode and its coupling strength. Here, we show that elements with even a relatively small footprint and, therefore, a small number of coupled TLS defects can significantly affect the quality factor of the qubit and have loss tangents close to those we obtained in this work. We simulate a qubit coupled with $i$ TLS defects with the strength $g_i$, having relaxation rate $\Gamma_{1TLS,i}$. We assume Markovian decoherence and calculate the qubit relaxation rate $\Gamma_1$[27,40]

$$\Gamma_1 = \sum_i \frac{2g_i^2 \Gamma_{1TLS,i}}{\Gamma_{1TLS,i}^2 + \Delta_i^2} \qquad (7)$$

where $\Delta_i$ is detuning. The model assumes the limit $\Gamma_{1TLS,i} > g_i$, where $\Gamma_{1TLS,i}$ is the defect relaxation rate. We simulate the distribution of electromagnetic field throughout the qubit using the same method as for calculating the participation ratios, applying the root mean square voltage to the qubit capacitive pads. We then sample random TLS defects from the dataset of Gaussian distributed dipole moments $d$ with mean dipole moment $d_{mean} = 2.6\ D$ and standard deviation $\sigma = 1.6\ D$ in accordance with the recent study[43]. We assign the decoherence rate to the defects, measured in Ref.[14,40,44–46], and a random detuning in the considered range of 300 MHz. We then place the sampled TLSs in the interface at a random position on the qubit to determine their coupling strength ($g = Ed$), and then using Eq. 7 calculate the spectrum of the qubit relaxation rate. To determine the amount of TLS, we took the previously obtained TLS density in amorphous alumina[35] $\rho_o \sim 1800\ (\mu m^3 \cdot GHz)^{-1}$ and set the effective thicknesses of the MA, MS, SA interfaces to 2, 0.3 and 0.36 nm, respectively. We assumed the MA interface thickness to be approximately equal to the native Al oxide interface thickness. The density of defects in the materials of the MS and SA interfaces is unknown, therefore the thicknesses of the MS and SA interfaces are set in accordance with the proportions of the interfaces loss tangents, experimentally extracted in Ref.[19] ($\tan\delta_{MA} = 3.9 \times 10^{-3}, \tan\delta_{MS} = 7.1 \times 10^{-4}, \tan\delta_{SA} = 5.9 \times 10^{-4}$), since the loss tangent scales with the defect density[47] $\tan\delta = \pi\rho_o d^2 (3\epsilon)^{-1}$. Note that the model according to Eq. 7 is used for strongly coupled defects, so we use it for the outer perimeter of capacitive pads and the ground plane, leads and SQUID, where the electric field is strong enough. For the inner areas of the pads with a weak electric field, we enter the background relaxation rate level, calculated using the Eq. 1 with known loss tangents[29]. After simulating the relaxation rate spectra of qubits with three lead designs ("long leads", "regular leads", "wide leads"), we apply the SLE process as in the experiment and extract the loss tangents of the qubit elements: $\tan\delta_{pads} = (8.4 \pm 4.4) \times 10^{-4}$, $\tan\delta_{leads} = (6.8 \pm 4.8) \times 10^{-4}$, $\tan\delta_{SQUID} = (3.2 \pm 2.8) \times 10^{-4}$.

The loss tangents obtained from TLS simulation are similar to the experimentally extracted ones. The proportions between the loss tangents of the elements also correspond to experimental data. A comparison of the proposed model and experimental data indicates that even a small fraction of TLS's located within the wiring interfaces of about 18% (total number of defects in the leads $N_{leads} \sim 4300/GHz$ and in the SQUID loop $N_{SQUID} \sim 950/GHz$) have a significant contribution to total qubit dielectric loss and limit qubit relaxation time. See Supplementary for simulated qubit quality factor spectra.

**Summary**

We have studied the wiring surface dielectric loss of superconducting transmon qubits and have found out that relaxation time of qubits with a large footprint is limited by dielectric loss in the leads. We considered three different transmon geometries and experimentally extracted the loss tangents of the capacitive pads, leads and SQUIDs using the SLE process. It is demonstrated, that for a commonly used tunable floating transmon qubit design, which we called "regular" in the work, leads and SQUIDs contribute about 50% of the total dielectric loss. We do not include variation of the capacitor pads participation ratios, which could improve the accuracy of extracted loss tangents, meanwhile, we confirmed, that leads can be a limiting factor for qubit relaxation times.

Wiring leads are often fabricated together with Josephson junctions using lift-off process, which introduce additional surface dielectric loss. We experimentally extracted the loss tangents for wet etched ($tg\delta_{leads} = (7.9 \pm 3.8) \times 10^{-4}$) and lift-off ($tg\delta_{leads} = (9.2 \pm 4.5) \times 10^{-4}$) leads. In order to minimize internal qubit surface loss and improve relaxation time, one should fabricate wiring leads together with capacitor pads using etching process. A further leads loss tangent reduction may be achieved by reducing their participation ratio with a substrate trenching[19,29,31,48].

We simulated qubit spectra by randomly sampling TLS in the qubit interfaces. The simulation shows that only a small fraction of coupled defects is located within the wiring interfaces (~ 18%), however, their coupling strength is much higher due to the stronger electric fields, which results in non-negligible loss of the wiring elements.

Finally, we demonstrated that electric field dilution by increasing the wiring width improves qubit performance up to 20% for the considered qubit designs. Further optimization of leads design by tapering[37], for example, may increase a relaxation time even more. In addition, one should pay attention to SQUID design, as it has a high participation ratio and loss tangent.

**METHODS**
**Sample fabrication**
The sample is fabricated on 525 $\mu m$ thick high-resistivity silicon. Firstly, the substrate is cleaned with a Piranha solution at 80 ℃, followed by dipping in 2% hydrofluoric bath. Then 120 nm aluminum film is deposited using e-beam evaporation in an ultrahigh vacuum deposition system. After that, 600-$nm$ thick positive photoresist is spin coated. Then ground plane, resonators and qubit capacitors are defined using a laser direct-writing lithography system. Josephson junction wiring of the qubits of the «wet etched» group are also patterned in this step. Patterned features are then wet etched using commercial Al etchant solution. The photoresist is stripped in N-methyl2-pyrrolidone at 80 ℃ for 3 hours and rinsed in IPA (isopropyl alcohol).

Josephson junctions, SQUID loops and wiring (for the «lift-off» group of the qubits) are then defined using Niemer-Dolan method. The process is described in details in Ref.[49,50]. The substrate is spin-coated with a resist bilayer composed of 500 nm MMA (methyl methacrylate) and 300 nm PMMA (poly methyl methacrylate). The development is performed in a bath of MIBK/IPA 1:3 solution followed by rinsing in IPA. Josephson junctions and wiring are patterned using an electron beam lithography system and then electrodes are shadow-evaporated in an ultra-high vacuum deposition system. First evaporated Al junction electrode is 25-nm thick and the second is 45-nm. Then aluminum bandages are defined and evaporated using the same process as for the junctions with an in-situ Ar ion milling. Lift-off is performed in a bath of N-methyl2-pyrrolidone with sonication at 80 ℃ for 3 h and rinsed in a bath of IPA with sonication.

Finally, aluminum free-standing crossovers are fabricated for the suppression of parasitic modes, using a common fabrication process. 3 $\mu m$ photoresist is spincoated and then the sacrificial layer is patterned using a direct laser writing system. A 300 nm of Al is then evaporated with an in-situ Ar ion milling to remove the native oxide. The second layer of 3 $\mu m$ photoresist is used as a protective mask and the excess metal is wet etched. A damaged layer of photoresist is then removed in oxygen plasma and both layers of photoresist are stripped of N-methyl2-pyrrolidone at 80 ℃.

**Experimental setup**

The detailed experimental setup scheme is shown in Fig. 4. The chip is connected to the control setup with seven lines: one line used both for readout and applying single qubit gates (XY controls) and six flux control lines used for detuning, coupled to each qubit. Pulsed XY control of the qubits was realized by upconverting the intermediate-frequency in-phase and quadrature signals from the arbitrary waveform generator (AWG), using IQ-mixer and microwave local oscillator. Detuning pulses were generated by single AWG channels. Readout tone was generated by AWG and up-converted to the readout resonator frequency using mixer and microwave local oscillator. The readout and XY control lines are combined by a 2-way splitter/divider. Readout microwave signal passes through the chip, is amplified by a cryogenic impedance matched parametric amplifier (IMPA)[51], and then downconverted. The readout signal is also amplified by high-electron mobility transistors (HEMT) at the 4K stage of the cryostat and at room temperature. We use a DC source and superconducting coil located on the packaging, to tune-up the parametric amplifier to the desired frequency and pump the IMPA by microwave source of vector network analyzer (VNA). Readout lines are additionally equipped with custom-made Eccosorb filters[39] on the cryostat mixing stage to suppress IR-noise and standing waves. Sample holders with qubit chips and IMPAs are both placed in the magnetic shields.

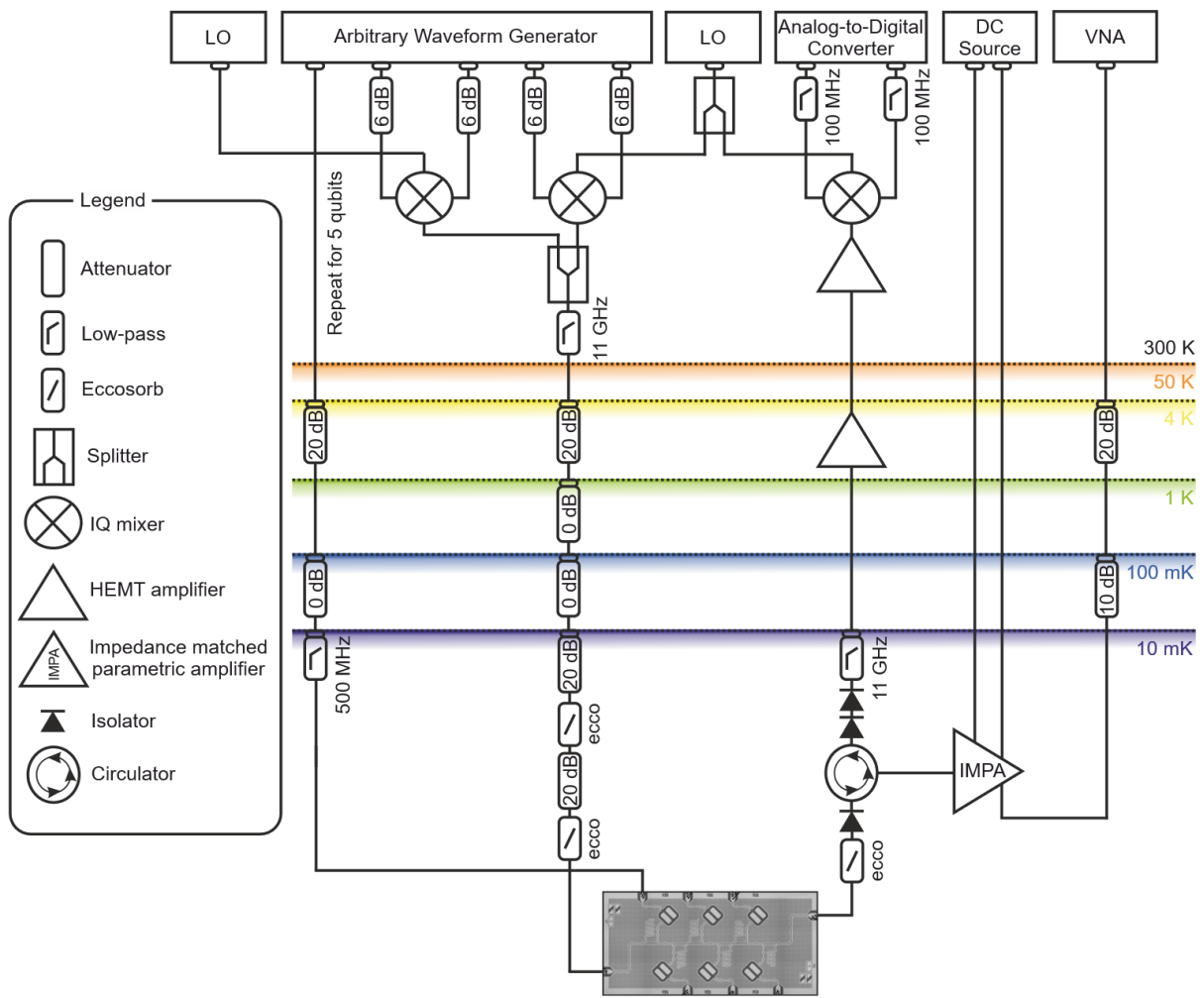

**Fig. 4 Schematic diagram of the experimental setup.**

## DATA AVAILABILITY

The data that support the findings of this study are available from the corresponding author upon reasonable request.

## ACKNOWLEDGEMENTS

Technology was developed and samples were made at the BMSTU Nanofabrication Facility (FMN Laboratory, FMNS REC, ID 74300).

## AUTHOR CONTRIBUTIONS

I.A.R. conceived the project and supervised the experiment. The experiments were devised by N.S.S. and I.A.R. The experimental and simulated data was analyzed by N.S.S. with E.A.K. FEM simulations were performed by E.A.K. The devices was designed by N.S.S. The device was fabricated by N.S.S and A.A.S. The measurements were performed by A.I.I. N.S.S. wrote the manuscript with I.A.R. and A.I.I. All authors discussed the results and the manuscript.

*Supplementary Material to*

# Wiring surface loss of a superconducting transmon qubit

**Nikita S. Smirnov,[1,2] Elizaveta A. Krivko,[1] Anastasiya A. Solovieva,[1] Anton I. Ivanov,[1] and Ilya A. Rodionov[1,2]***

[1]FMN Laboratory, Bauman Moscow State Technical University, Moscow, 105005, Russia

[2]Dukhov Automatics Research Institute, VNIIA, Moscow, 127030, Russia

* irodionov@bmstu.ru

## DETAILS ON SURFACE PARTICIPATION SIMULATION

The use of participation ratios is convenient for estimating dissipative losses in systems with dielectrics. The method for calculating qubit participation ratios used in the paper is based on the Ref S1 with some modifications. In this method, calculations are carried out in two steps: a 3D simulation of the entire qubit at a coarse scale (~ μm) (Fig. S1a) and fine simulation of qubit regions with highly concentrated electric fields at the edges of the electrode pads and narrow Josephson junction wiring. Dividing the simulation into stages helps to overcome the problem of obtaining convergent values of electric fields in thin dielectric layers, caused by the large difference between the length scales of the system.

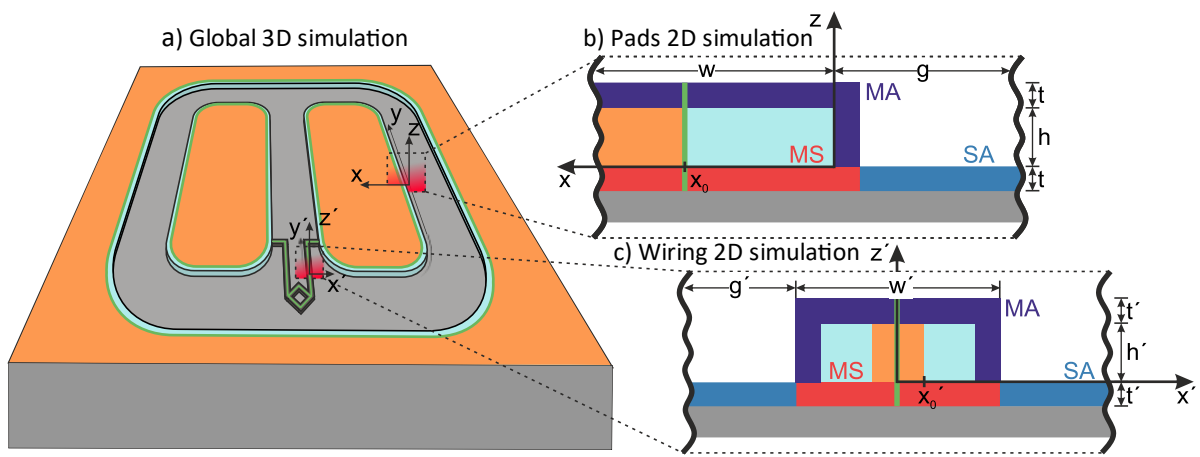

**Fig. S1 Illustration of the two-step computation of transmon qubit surface participation ratios. a** The global coarse 3D simulation which includes features of the entire qubit, such as substrate, capacitor pads, ground plane, leads and SQUID loop. Conductors are perfectly-conducting sheets, interface layers are omitted. The inner (orange) and perimeter (blue) regions are separated by the green contour. The red windows represent the placements of the cross-sections. **b** Cross-section of the capacitor pad edge used for 2D electrostatic simulation. Interfaces with thickness of *t* are shown in red (MS), purple (MA) and blue (SA). Superconductor regions with thickness of *h* are represented in orange that is inner region and light blue that is perimeter region. Ground plane edge has the same cross-section. **c** Cross-section of the wiring, used for 2D simulation of wiring, including long leads and a SQUID loop. All color codes are the same as for the cross-section of the capacitor pad edge.

First, the electrode pads and ground plane are simulated (Fig.S1b). Electrode pads are divided into «perimeter regions» and «inner regions» with a boundary set at a distance $x_0$ from the edge equal to 1 $\mu m$. Electric field of the inner regions easily converges in 3D simulations since it is located far away from the edge of the electrodes and uniformly distributed. From these regions we immediately extract the electric field values on the top and bottom surfaces of the electrode pads and ground plane, surrounding qubits ($E_{MA}(x,y)$ and $E_{MS}(x,y)$), as well as on the surface of exposed substrate areas $E_{SA}(x,y)$. The field distributions are then used to calculate the inner participation ratios of the regions associated with the capacitor pads, substrate and ground plane:

$$p_{i,int} = t \iint_{int} \frac{\epsilon}{2} |E_i(x,y)|^2 dxdy / U_{tot} \qquad (S1)$$

where $i$ is MA, MS or SA, $U_{tot}$ is the total electric field energy in the entire model, $t$ is the thickness of the dielectric layer, which we assumed to be equal to 3 nm in our calculations. Figure S2 illustrates the results of coarse electric field simulation.

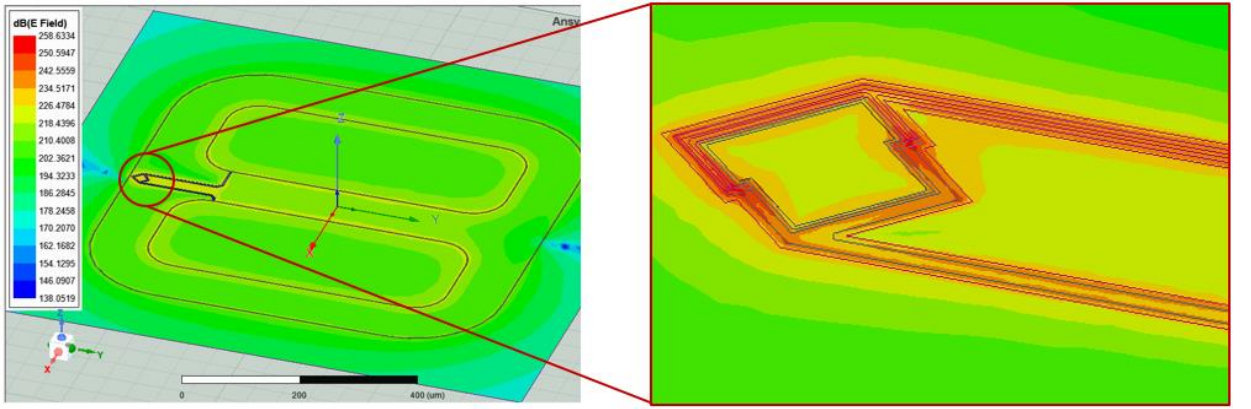

**Fig. S2 Resulting electric filed distribution from coarse global simulation.** The electric field is highly concentrated around the Josephson junction wiring.

At the second stage, the perimeter regions are calculated. Here, in a 2D simulation, a constant ratio $F$ of the integrated field energy within the entire cross-section of the perimeter region ($0 < x < x_0$) and the section, that converges both in the coarse 3D simulation, and the 2D simulation ($x_0/2 < x < x_0$). This constant scaling factor is then used to calculate participation ratios of the entire perimeter region:

$$p_{i,per} = F_i t \int_{x_0/2}^{x_0} dx \oint_y \frac{\epsilon}{2} |E_i(x,y)|^2 dy / U_{tot} \qquad (S2)$$

Electric field distribution in the volume at the edges of the electrode $f(x,z)$ is independent of the pad width, gap width and the other distant boundary conditions so that $E(x,y,z) = C(y)f(x,z)$. Figure S3a shows spatial distribution of the electric field $f(x,z)$ for different pad width $w$ and gap width $g$, including typical values, used in this study. Using the simulated $f(x,z)$ scaling factors $F_i$ for the electrode pads within 2D cross-section are simulated:

$$F_{MS} = \frac{\int_0^{x_0} dx \int_{-t}^0 f^2(x,z)dz}{\int_{x_0/2}^{x_0} dx \int_{-t}^0 f^2(x,z)dz} \qquad (S3)$$

$$F_{MA} = \frac{\int_0^{x_0} dx \int_h^{h+t} f^2(x,z)dz + \int_{-t}^0 dx \int_0^{h+t} f^2(x,z)dz}{\int_{x_0/2}^{x_0} dx \int_h^{h+t} f^2(x,z)dz} \qquad (S4)$$

$$F_{SA} = \frac{\int_0^{-x_0} dx \int_{-t}^0 f^2(x,z)dz}{\int_{x_0/2}^{x_0} dx \int_{-t}^0 f^2(x,z)dz} \tag{S5}$$

Figure S3b shows scaling factor of the MS interface as a function of $x_0$ position for different geometries.

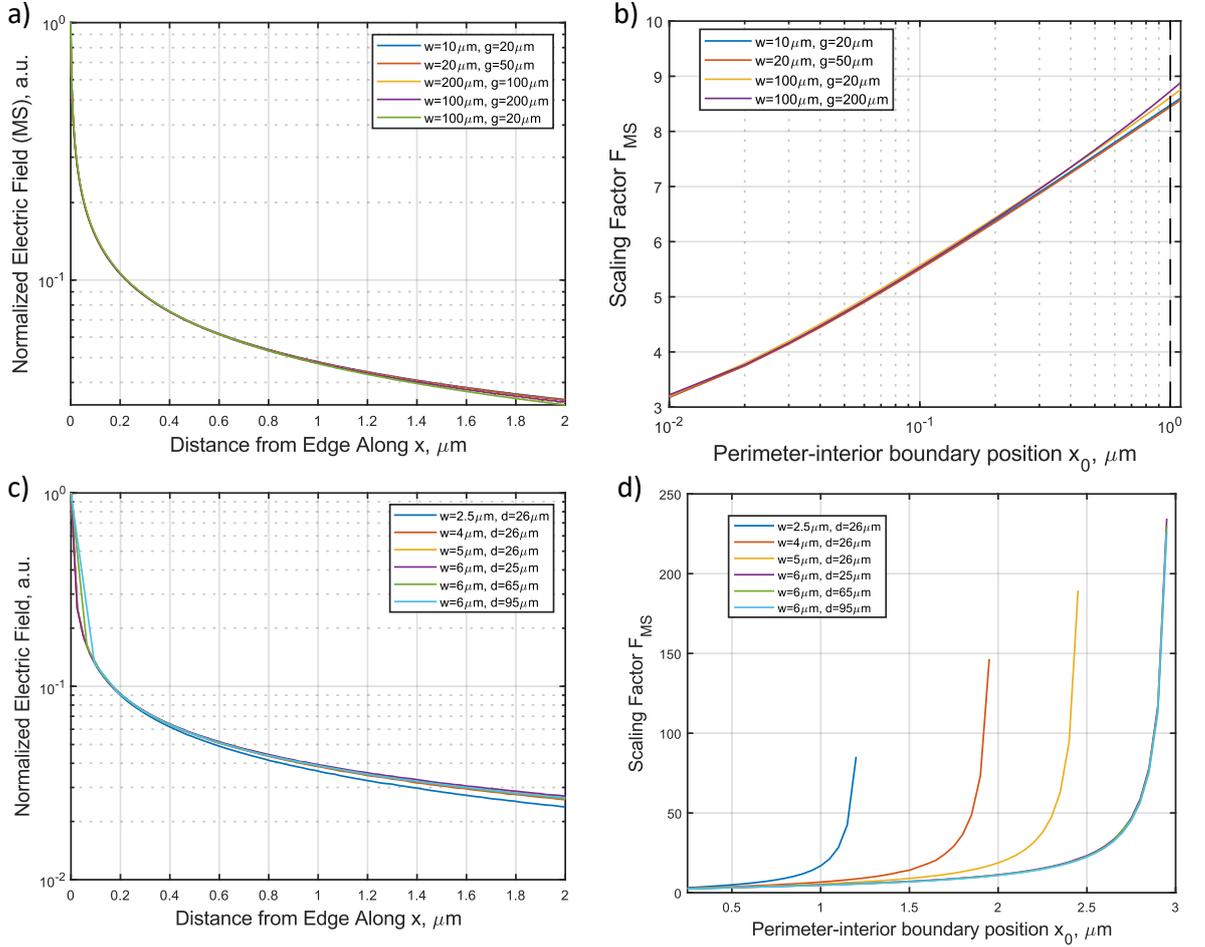

**Fig. S3 Results of cross-sectional electric field simulations.** Values are plotted for different distant boundary conditions. $w$ represents width of the conducting feature, $g$ and $d$ represent distance between features for capacitor edge and leads respectively. **a** Normalized electric field distribution within a cross section of the MS interface near a capacitor edge. **b** $F_{MS}$ scaling factors as function of $x_0$ position plotted for various boundary conditions. **c** Normalized electric field distribution within a cross section of the SA interface between the wiring features near the lead edge. **d** $F_{MS}$ scaling factors as function of $x_0'$ position plotted for various wiring geometries. $F_{MS}$ is the same for all varied boundary conditions when $x_0' = 0.5\ \mu m$.

The participation ratios of the qubit wiring are calculated similarly to the capacitor pads (Fig S1c). Since in our qubit geometry the wiring leads are parallel to each other, we can, as for the capacitor pads, find a constant ratio $F'$ in 2D simulation, and then using the field distribution from the global coarse 3D simulation, calculate the participation ratios. The spatial field distribution at the edges of the leads $f'(x,z)$ is independent of the lead width and distance between the leads for the wiring geometries used in our study. Figure S3c shows normalized field distributions $f'(x,z)$ within SA interface for different leads width $w'$ and distance between the leads $g'$. In 2D simulation we consider scaling factor $F'$ of the entire leads cross-section ($-w'/2 < x < w'/2$) and the section, that converges in both coarse global simulation and fine 2D simulation ($-x_0' < x < x_0'$). The distance $x_0'$ within which electric fields converge in both types of simulation is set to be equal 0.5 $\mu m$. For better convergence we set the finer meshing on the leads than on the capacitor pads. Scaling factor is then used to calculate participation ratios of the wiring.

$$p_{i,leads} = F_i' t' \int_{-x_0'}^{x_0'} dx' \oint_{y'} \frac{\epsilon}{2} |\mathbf{E}_i(x',y')|^2 \, dy' / U_{tot} \tag{S6}$$

Scaling factors $F_i'$ within the wiring cross-section are calculated as follows:

$$F_{MS}' = \frac{\int_{-w'/2}^{w'/2} dx' \int_{-t'}^{0} f^2(x',z') dz'}{\int_{-x_0'}^{x_0'} dx' \int_{-t'}^{0} f^2(x',z') dz'} \tag{S7}$$

$$F_{MA}' = \frac{\int_{-\frac{w'}{2}+t'}^{\frac{w'}{2}-t'} dx' \int_{h'}^{h'+t'} f^2(x',z') dz' + \int_{-\frac{w'}{2}}^{-\frac{w'}{2}+t'} dx' \int_{0}^{h'+t'} f^2(x',z') dz' + \int_{\frac{w'}{2}-t'}^{\frac{w'}{2}} dx' \int_{0}^{h'+t'} f^2(x',z') dz'}{\int_{-x_0'}^{x_0'} dx' \int_{h'}^{h'+t'} f^2(x',z') dz'} \tag{S8}$$

$$F_{SA}' = \frac{\int_{-\frac{w'}{2}-2x_0}^{-\frac{w'}{2}} dx' \int_{-t'}^{0} f^2(x',z') dz' + \int_{\frac{w'}{2}}^{\frac{w'}{2}+2x_0} dx' \int_{-t'}^{0} f^2(x',z') dz'}{\int_{-x_0'}^{x_0'} dx' \int_{-t'}^{0} f^2(x',z') dz'} \tag{S9}$$

Figure S3d shows scaling factor of the MS interface as a function of $x_0'$ position for different wiring cross-sections. The calculated participation ratios of the qubits used in this study are presented in Table S1.

**Table S1. Results of participation ratios for the transmon qubits used in the paper.**

| Qubit# (etch/lift-off) | Design | $p_{pads_{MA}}$ | $p_{pads_{MS}}$ | $p_{pads_{SA}}$ | $p_{leads_{MA}}$ | $p_{leads_{MS}}$ | $p_{leads_{SA}}$ | $p_{SQUID_{MA}}$ | $p_{SQUID_{MS}}$ | $p_{SQUID_{SA}}$ |
|---|---|---|---|---|---|---|---|---|---|---|
| 1/6 | long | 0.0362 | 0.5435 | 1.2723 | 0.1905 | 1.7733 | 1.3484 | 0.0353 | 0.3282 | 0.2496 |
| 3/4 | regular | 0.0487 | 0.7033 | 1.1862 | 0.0718 | 0.6679 | 0.5078 | 0.0399 | 0.3714 | 0.2824 |
| 5/2 | wide | 0.0170 | 0.4105 | 1.6588 | 0.0359 | 0.3605 | 0.2555 | 0.0398 | 0.4003 | 0.2837 |

Participation ratios values in the table are multiplied by $10^4$.

## DETAILS ON MONTE CARLO SIMULATION OF QUBIT RELAXATION RATE SPECTRUM

To compare the experimental results with the TLS model, we performed a simulation of a qubit coupled to a TLS bath, as described in the paper. We simulated three qubit geometries, the same as those used in our study. We used the TLS parameters that were experimentally extracted in the previous studies and simulated the electric field to calculate the qubit-defect coupling strength. Simulated spectra and Q-factor histograms of the qubits converted from relaxation rates and rescaled to frequencies from 4 to 5 GHz are presented in Fig. S4a at 1 GHz bandwidth.

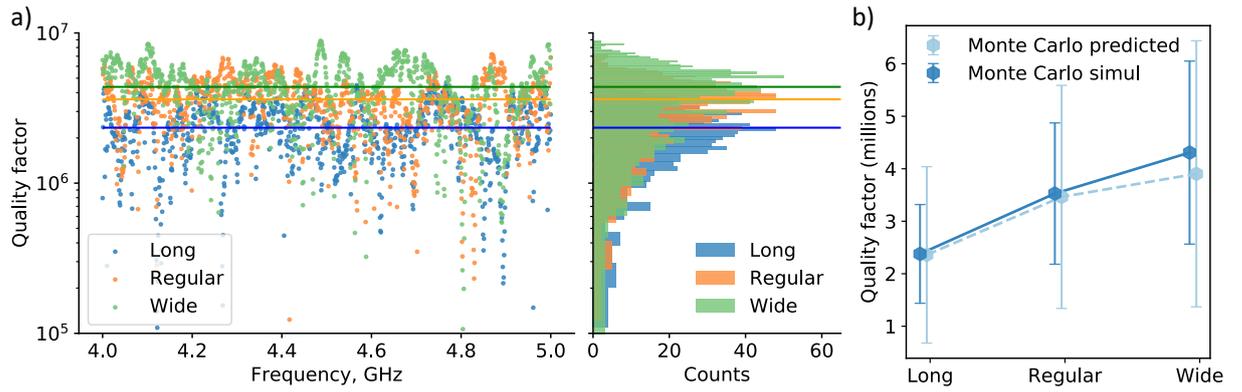

**Fig. S4 Monte Carlo simulation of defects in the qubits. a** Simulated Q-factor spectra and distribution. Blue, orange and green data correspond to the "long leads", "regular leads" and "wide leads" respectively. Solid horizontal lines show the median Q-factor values of the corresponding qubit. **b** Monte Carlo simulated and predicted quality Q-factors as a function of leads geometry. Each data point represents median Q-factor obtained from the simulated qubit Q-factor spectra or sampled Q-factors using the SLE process. Error bars represent 68% confidence interval. Dark blue data points represent Q-factors from the simulated spectra. Light blue data points represent predicted Q-factors.

We performed the SLE process on the data obtained from Monte Carlo simulations of qubit spectra. Extracted loss tangents are presented in the main text. The calculated quality factors from the extracted loss tangents vs. the simulated Q-factors are shown as a dashed line in Fig. S4b. Error bars represent 68% confidence interval. The shape of the predicted curve matches the simulated data and is similar to the curves obtained from the experiment.